\documentclass[superscriptaddress, 
floatfix,showkeys,nofootinbib]{revtex4-2} 
\usepackage{times,fancyhdr}
\usepackage[dvips]{graphicx}
\usepackage{amsmath,amssymb,bm}
\usepackage{color}
\usepackage{mathrsfs}
\usepackage{setspace}
\usepackage{hyperref}
\usepackage{natbib}

\def\beq{\begin{equation}}
\def\eeq{\end{equation}}
\def\beqn{\begin{align}}
\def\eeqn{\end{align}}

\begin{document}
\title{Parametric factorization of non linear second order differential equations}
\author{Gabriel Gonz\'alez}
\affiliation{C\'atedra CONAHCYT--Universidad Aut\'onoma de San Luis Potos\'i, San Luis Potos\'i, 78000 MEXICO}
\affiliation{Coordinaci\'on para la Innovaci\'on y la Aplicaci\'on de la Ciencia y la Tecnolog\'ia, Universidad Aut\'onoma de San Luis Potos\'i,San Luis Potos\'i, 78000 MEXICO}

%

\begin{abstract}
In this paper the factorization method introduced by Rosu \& Cornejo-P\'erez for second order non linear differential equations is generalized by adding a parameter in order to obtain the general solutions for the mixed quadratic and linear Li\'enard type equation. The new parametric factorization is used to obtain complete analytic solutions for nonlinear second order differential equations. The parametric factorization introduced in this article reduces to the standard factorization scheme when the parameter goes to zero. As an example, we apply the parametric factorization approach to solve the generalized Fisher equation and the Israel-Stewart cosmological model. The parametric factorization presented in this paper can be used in other non linear mixed Li\'enard type equations.

\end{abstract}

\maketitle

\section{Introduction}
Non linear second order differential equations are widely used to describe various phenomena in physics and mathematics and the vast majority of them do not have analytic solutions and are very difficult to analyze them. Developing methods for finding solutions for non linear differential equations has been a problem of interest for a long time. At the present time there are many methods for finding exact solutions of non linear equations such as: the $\tanh$-expansion method \cite{wazwaz2007tanh}, the Darboux transformation \cite{matveev1991darboux}, the B\"acklund transformation \cite{levi1981nonlinear}, Hirota bilinear method \cite{hietarinta2007introduction}, Painlev\'e truncation expansion \cite{lou1998extended,costin2018truncated},  generalized Sundman transformation \cite{nakpim2010linearization}, point transformations and contact transformations\cite{levinson1944transformation}. All these approaches have yielded many interesting exact solutions of the kink and soliton type for well-known nonlinear equations. \\
Despite the vast and rich variety of different methods to solve non linear differential equations, the fundamental problem of finding explicit and exact analytic solutions to nonlinear differential equations constitutes yet an active area of research. In the process of learning to solve non linear differential equations it is convenient to begin with simple and efficient methods which provide a way to obtain exact analytic solutions. One of the most popular existing simple methods to solve non linear differential equations is by using travelling wave transformations and direct integration. However, the travelling wave transformation method is only suitable for a certain type of non linear differential equations. Recently, Rosu \& Cornejo proposed a factorization method which allows one to obtain travelling wave solutions of the reaction-diffusion equations with polynomial nonlinearities\cite{rosu2005supersymmetric}. Using the factorization method, Cornejo \& Rosu obtained particular solutions of several important equations, among which are the Fisher equation, the FitzHugh-Nagumo equation and the generalized Burgers-Huxley equation, which maps into a second order non linear differential equation of Li\'enard type in the travelling coordinate frame of the form\cite{cornejo2005nonlinear}
\begin{equation}\label{eqi1}
  \ddot{x}+f(x)\dot{x}+g(x)=0
\end{equation}
where $f(x)$ and $g(x)$ are polynomial functions and overdot denotes differentiation with respect to time. \\
In this paper we extend the factorization method introduced by Rosu \& Cornejo to solve a mixed quadratic-linear Li\'enard type equation of the form
\begin{equation}\label{eqi2}
  \ddot{x}+\mu\frac{\dot{x}^2}{x}+F(x)\dot{x}+G(x)=0
\end{equation}
where $\mu$ is an auxiliary parameter to be determined. One can see that equation (\ref{eqi1}) is a subcase of equation (\ref{eqi2}). In applications one often encounters differential equations in which both linear and quadratic terms are present. Equation (\ref{eqi2}) has a particular form of a mixed type Li\'enard equation which frequently appears as a mathematical model in several areas of physics, for example equation (\ref{eqi2}) belongs to the type of second order Gambier equation when the coefficients are assumed to be constant parameters. The Gambier equation written as a second order differential equation takes the form\cite{carinena2013quasi}
\begin{align}\label{eqi3}
  \ddot{x}=&\frac{n-1}{n}\frac{\dot{x}^2}{x}+\left(\frac{n+2}{n}ax-\frac{n-2}{n}\frac{\sigma}{x}+b\right)\dot{x} \\
   &-\frac{a^2}{n}x^3+(\dot{a}-ab)x^2+\left(cn-\frac{2a\sigma}{n}\right)x-b\sigma-\frac{\sigma^2}{nx}.
\end{align}
where $a$, $b$ and $c$ are functions of the independent variable and $\sigma$ is a constant. Interestingly, the parametric factorization introduced in this paper is of the type of an autonomous second order Gambier equation. The importance of the second order Gambier equation is due to the fact that it is related with very important non linear differential equations, such as the second order Ricatti equations, second order Kummer-Schwartz equation and Milne-Pinney equation, to name a few\cite{pradeep2010certain}. More recently, Zheng and Shang\cite{zheng2015abundant} showed that the amplitude part of the solution in phase amplitude format of the nonlinear Schr\"odinger equation with dual power nonlinearities satisfies a mixed Li\'enard type equation of the form given in equation (\ref{eqi2}). Thus, the goal of this paper is to present an extension of the factorization method in which one can obtain solutions of a mixed Li\'enard type equation. The parametric factorization introduced in this paper has the advantage that allows to obtain solutions of second order non linear differential equations with linear and quadratic damping terms and contains as a particular case the standard factorization.   \\
The article is organized as follows. In the first section we will review the Rosu \& Cornejo-P\'erez factorization scheme and introduce the parametric factorization. In the second section we will apply the parametric factorization to obtain particular and parametric solutions of the generalized Fisher. In the third section we will use the parametric factorization to obtain particular and parametric solutions for the Israel-Stewart cosmological model. The conclusions are summarized in the last section.
\section{Parametric Factorization}
\begin{figure}[ht]
  \centering
  \includegraphics[height=5cm]{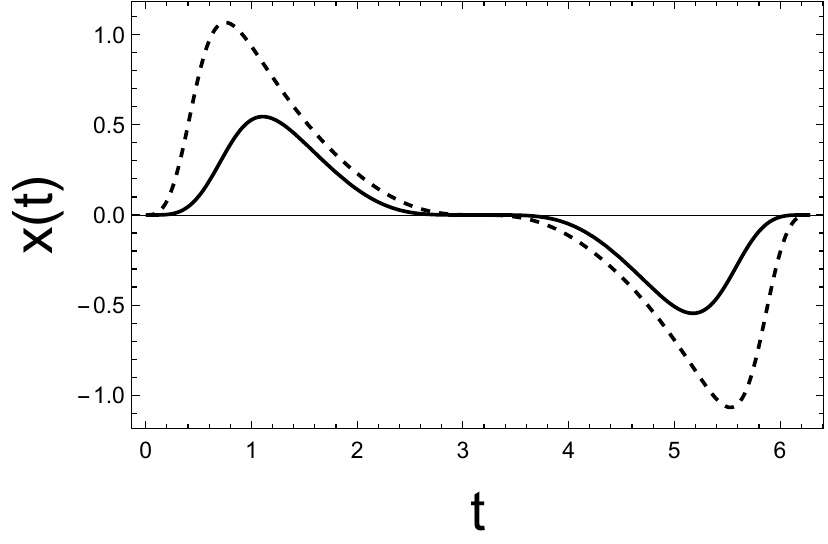}
   \caption{The figure shows the solutions $x(t)$ of the mixed Li\'enard type equation for $m=1$ and $m=2$ respectively.}
  \label{fig1}
\end{figure}

An elegant procedure to solve second order non linear differential equations consists in using the factorization method, where a given non linear differential operator is factorized in two first order differential operators.
In 2005, Rosu and Cornejo-P\'erez \cite{rosu2005supersymmetric} introduced an effective factorization of second-order ordinary differential equations with polynomial nonlinearities by taking additional advantage from the polynomial factorization of the nonlinear part. Using the factorization technique, Rosu and Cornejo-P\'erez obtained particular solutions of the following Li\'enard type differential equation
\begin{equation}\label{eq0}
  \ddot{q}+f(q)\dot{q}+g(q)=0,
\end{equation}
where the dot represents the derivative with respect to time. Equation (\ref{eq0}) admits the following factorization
\beq\label{eq2}
\left(D-\phi_2(q)\right)\left(D-\phi_1(q)\right)q=0~,\qquad D=\frac{d}{dt}~.
\eeq
By expanding equation (\ref{eq2}) and comparing with equation (\ref{eq0}), one obtains the following conditions over the functions $\phi_1$ and $\phi_2$:
\begin{align}\label{eqc0}
&\phi_1+\phi_2+q\frac{d\phi_1}{dq}=-f(q) \\
&\phi_1\phi_2=\frac{g(q)}{q}~.
\end{align}
To obtain a particular solution, Rosu and Cornejo-P\'erez solved the following first order differential equation
\begin{equation}\label{eq3}
  \left(D-\phi_1(q)\right)q=0
\end{equation}
obtaining a particular solution of (\ref{eq1}) by one quadrature
\begin{equation}\label{eq4}
t-t_0=\int \frac{dq}{q \phi_1(q)}~
\end{equation}
In this article we will extend the factorization method outlined above by adding a parameter $\mu$ in equation (\ref{eq2}) such that the factorization is given now by
\begin{equation}\label{eq1}
\left(D-\varphi_2(x)\right)\left(D-\varphi_1(x)\right)x^{\mu +1}=0,
\end{equation}
where $\mu\neq-1$.
If we expand the factorization given in equation (\ref{eq1}) we get the following non linear second order differential equation
\begin{equation}\label{eq5}
  \ddot{x}+\mu\frac{\dot{x}^2}{x}-\dot{x}\left(\varphi_1(x)+\varphi_2(x)+\frac{x}{\mu+1}\frac{d\varphi_1}{dx}\right)+\frac{x}{\mu+1}\varphi_1(x)\varphi_2(x)=0.
\end{equation}
Equation (\ref{eq5}) is a particular form of the mixed Li\'enard type equation with quadratic and linear terms. Note that equation (\ref{eq5}) reduces to the standard Li\'enard type equation when $\mu\rightarrow 0$. By comparing (\ref{eqi2}) and (\ref{eq5}), one obtains the conditions for the parametric factorization over the functions $\varphi_1$ and $\varphi_2$:
\begin{align}
&\varphi_1+\varphi_2+\frac{x}{\mu+1}\frac{d\varphi_1}{dx}=-F(x) \label{eqc} \\
&\varphi_1\varphi_2=(\mu+1)\frac{G(x)}{x}~. \label{eqc1}
\end{align}
An interesting feature between the standard factorization and the parametric factorization is that we can transform one to the other by a non trivial space transformation once written in factorized form. This result allows us to map solutions into solutions between the Li\'enard equation (\ref{eqi1}) and the mixed Li\'enard equation (\ref{eqi2}). Let us now work out a simple example to illustrate this point. Consider the following Li\'enard type equation given by
\begin{equation}\label{eq5a}
  \ddot{q}+(2m+3)q^{2m+1}\dot{q}+q+q^{4m+3}=0,
\end{equation}
with $m$ a non negative integer. Equation (\ref{eq5a}) represents a class of solvable nonlinear oscillators with isochronous orbits \cite{iacono2011class}, i.e. orbits with fixed period, not dependent with the amplitude. Equation (\ref{eq5a}) can be written in the following standard factorization form
\begin{equation}\label{eq5b}
  \left(D+q^{2m+1}+i\right) \left(D+q^{2m+1}-i\right)q=0
\end{equation}
We can make the following space transformation $q^{2m+1}=x$ so that equation (\ref{eq5b}) becomes
\begin{equation}\label{eq5c}
 \left(D+x+i\right) \left(D+x-i\right)x^{1/(2m+1)}=0
\end{equation}
Equation (\ref{eq5c}) is now written in the parametric factorization form where $\mu=-2m/(2m+1)$.  We can expand equation (\ref{eq5c}) to get
\begin{equation}\label{eq5d}
  \ddot{x}-\left(\frac{2m}{2m+1}\right)\frac{\dot{x}^2}{x}+(2m+3)x\dot{x}+(2m+1)(x+x^{3})=0.
\end{equation}
Equations (\ref{eq5b}) and (\ref{eq5c}) share the same solution given by
\begin{equation}\label{eq5e}
  q(t)=x^{1/(2m+1)}(t)=\frac{\sin(t-t_0)}{\left(C+(2m+1)\int\sin^{2m+1}(t-t_0)dt\right)^{1/(2m+1)}}
\end{equation}
where $t_0$ and $C$ are arbitrary constant. Therefore, the solution for the mixed Li\'enard equation (\ref{eq5d}) is given by
\begin{equation}\label{eq5f}
  x(t)=\frac{\sin^{2m+1}(t-t_0)}{C-\cos(t-t_0)\sum_{r=0}^{m}A_{mr}\sin^{2r}(t-t_0)}
\end{equation}
with
\begin{equation}\label{eq5g}
  A_{mr}=\frac{2^{2(m-r)}(m!)^2(2r)!}{(2m)!(r!)^2},
\end{equation}
where the condition for periodic solutions is $|C|>A_{m0}$ \cite{iacono2011class}. The solution for the mixed Li\'enard type equation given in equation (\ref{eq5d}) are shown in figure (\ref{fig1}) for $m=1$ and $m=2$, respectively.\\
Consequently, solutions from the standard factorization scheme and their properties can be used to obtain solutions of the parametric factorization approach through a space coordinate transformation.
\section{Generalized Fisher equation}
\begin{figure}[ht]
  \centering
  \includegraphics[height=5cm]{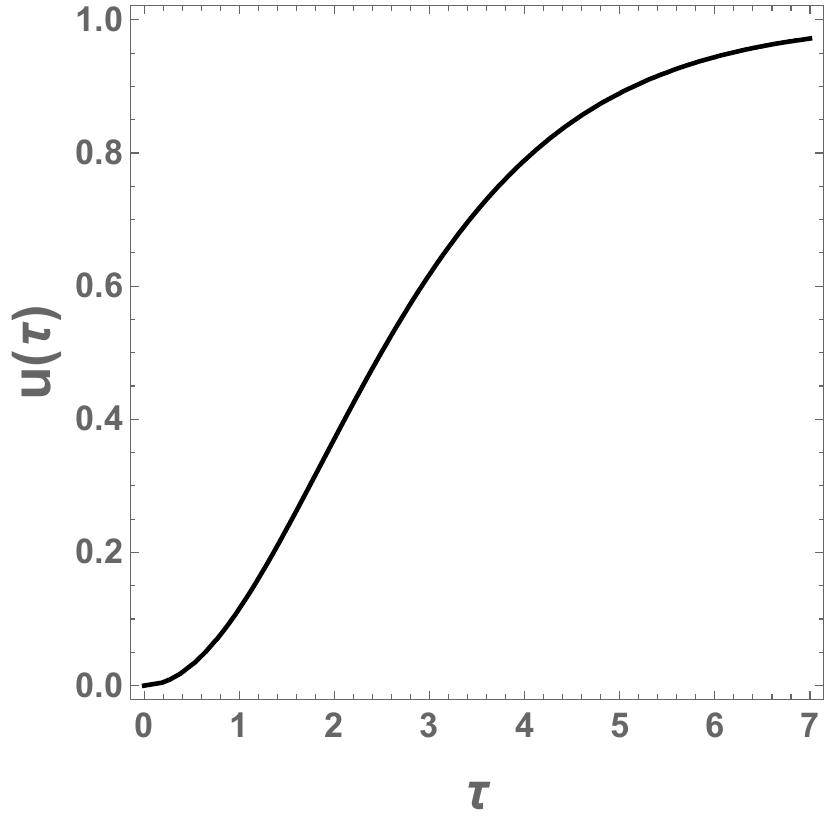}
   \caption{The figure shows the travelling wave solution $u(\tau)$ of the generalized Fisher equation for $m=1/4$, $p=3/4$ and $q=2$, moving with constant velocity of $v=3/\sqrt{2}$.}
  \label{fig2}
\end{figure}

We will now show how to apply the parametric factorization approach to obtain solutions of the generalized Fisher equation which is used in biology\cite{murray2002mathematical}. The generalized Fisher equation is a non linear partial differential equation which describes diffusion models for insects and biology invasion. From the perspective of biology invasion, the generalized Fisher equation predicts how the population of a particular species will spread via travelling waves. Let us consider the generalized Fisher equation given by\cite{kudryashov2014note}
\begin{equation}\label{eq6}
  \frac{\partial u}{\partial t}=u^p\left(1-u^q\right)+\frac{\partial}{\partial x}\left(u^m\frac{\partial u}{\partial x}\right)
\end{equation}
where $u$ represents the population density and $p$, $q$ and $m$ are positive parameters. Solutions for equation (\ref{eq6}) have been found for some values of the parameters $p$, $q$ and $m$; in particular, the standard factorization method has been used for the case $m=0$, $p=q=1$, i.e. the standard Fisher equation, and for the case $m=0$, $p=1$ and $q=2$, i.e. the Burguers-Huxley equation, respectively\cite{cornejo2005nonlinear}. In this section we will consider the generalized Fisher equation for the case $m\neq 0$ which can be written in the travelling reference frame $\tau=kx-\omega t=k(x-vt)$ as the following mixed Li\'enard non linear differential equation
\begin{equation}\label{eq7}
  \ddot{u}+m\frac{\dot{u}^2}{u}+\omega\frac{\dot{u}}{k^2u^m}+\frac{u^{p-m}}{k^2}-\frac{u^{p+q-m}}{k^2}=0
\end{equation}
where the over-dot represents $D=\frac{d}{d\tau}$. Using the second factorization condition given in equation (\ref{eqc1}) we have
\begin{equation}\label{eq8}
  \varphi_1\varphi_2=\frac{1+m}{k^2}u^{p-(1+m)}\left(1-u^q\right).
\end{equation}
Therefore, we can choose
\begin{eqnarray*}
  \varphi_1 &=& a_1\frac{\sqrt{1+m}}{k}u^{(p-(1+m))/2}\left(1-u^{q/2}\right) \\
  \varphi_2 &=&\frac{\sqrt{1+m}}{a_1k}u^{(p-(1+m))/2}\left(1+u^{q/2}\right)
\end{eqnarray*}
 where $a_1$ is a nonzero constant to be determined with the first factorization condition given in equation (\ref{eqc}) which reads
 \begin{align}\label{eq9}
  \varphi_1+\varphi_2+\frac{u}{\mu+1}\frac{d\varphi_1}{du}&=\frac{a_1}{k}u^{(p-(1+m))/2}\left[\sqrt{1+m}+\frac{p-(1+m)}{2\sqrt{1+m}}-u^{q/2}\left(\sqrt{1+m}+\frac{q-p+(1+m)}{2\sqrt{1+m}}\right)\right]\\ \nonumber
  &+\frac{\sqrt{1+m}}{a_1k}u^{(p-(1+m))/2}\left[1+u^{q/2}\right]=-\frac{\omega}{k^2u^m}
 \end{align}
 It follows that we have to choose $p=1-m$ in order to satisfy the parametric factorization conditions such that we get the following values for $a_1$ and $\omega$
 \begin{equation}\label{eq10}
   a_1=\pm\sqrt{\frac{2(1+m)}{2+q}}\quad \mbox{and}\quad \omega=\mp\frac{k(q+4)}{\sqrt{2(q+2)}}.
 \end{equation}
 Therefore, the travelling wave solutions we are going to obtain are moving with a constant velocity of $v=\omega/k=\mp\frac{(q+4)}{\sqrt{2(q+2)}}$, which means that with increasing value of $q$ the velocity modulus increases from $2$ to $\infty$.\\
 Equation (\ref{eq7}) admits the following parametric factorization
 \begin{equation}\label{11}
   \left[D \mp \sqrt{\frac{2+q}{2}}\frac{1}{ku^{2m}}\left(1+u^{q/2}\right)\right]\left[D\mp(1+m)\sqrt{\frac{2}{2+q}}\frac{1}{ku^{2m}}\left(1-u^{q/2}\right)\right]u^{m +1}=0.
    \end{equation}
 If one wants to find a particular solution to equation (\ref{eq7}) we have to solve only a compatible first order differential equation given by
 \begin{equation}\label{eq12}
   \left[D\mp(1+m)\sqrt{\frac{2}{2+q}}\frac{1}{ku^{2m}}\left(1-u^{q/2}\right)\right]u^{m +1}=0,
 \end{equation}
 which has the following implicit solution
 \begin{equation}\label{eq13}
   u^{2m}{_2}F_1\left[1,\frac{4m}{q},1+\frac{4m}{q},u^{q/2}\right]=\pm 2m\sqrt{\frac{2}{2+q}}(\tau-\tau_0)
 \end{equation}
 where $_2F_1$ is the hypergeometric function and $\tau_0$ is an integration constant.  In figure (\ref{fig2}) we show a travelling wave solution for the generalized Fisher equation obtained with the parametric factorization approach.\\
It is possible to solve the generalized Fisher equation given by
 \begin{equation}\label{eq14}
    \frac{\partial u}{\partial t}=u^{1-m}\left(1-u^q\right)+\frac{\partial}{\partial x}\left(u^m\frac{\partial u}{\partial x}\right)
 \end{equation}
 in a different way by using the parametric factorization conditions given in equations (\ref{eqc}) and (\ref{eqc1}) \cite{wang2008single}. Solving $\varphi_2$ from the first equation and substituting in the second equation we have
 \begin{equation}\label{eq15}
   -F(u)\varphi_1-\varphi_1\left(\varphi_1+\frac{u}{\mu+1}\frac{d\varphi_1}{du}\right)=(\mu+1)\frac{G(u)}{u},
    \end{equation}
which is transformed into an Abel equation of the second kind
\begin{equation}\label{eq16a}
  ww^{\prime}=F(u)u^{\mu}w-G(u)u^{2\mu}
\end{equation}
 by using the substitution
\begin{equation}\label{eq16}
  w(u)=-\frac{u^{\mu+1}\varphi_1(u)}{\mu+1}.
\end{equation}
The Abel equation given in (\ref{eq16a}) admits exact parametric solutions for special cases. For our particular case, we have $\mu=m$, $F(u)=\omega u^{-m}/k^2$ and $G(u)=u^{p-m}(1-u^{q})/k^2$, therefore we need to solve the following Abel equation given by
\begin{equation}\label{eq17}
  ww^{\prime}-\frac{\omega}{k^2}w=-\frac{u^{p-m}}{k^2}+\frac{u^{p-m+q}}{k^2}
\end{equation}
which for the case when $k^{-2}=2(2+q)/(4+q)^2$ and $\omega=(4+q)^2/(4+2q)$ has the following solution in parametric form\cite{zaitsev2002handbook}
\begin{equation}\label{eq18}
  u=\frac{(q+4)}{q}a\xi E^{2/q}_{1+q},\quad \mbox{and}\quad w=aE^{2/q}_{1+q}\left(R_{1+q}E_{1+q}+\frac{2}{q}\xi\right)
\end{equation}
where
\begin{equation}\label{eq19}
  E_{1+q}=\int(1\pm\xi^{2+q})^{-1/2}d\xi +C, \quad R_{1+q}=\sqrt{1\pm\xi^{2+q}} \quad \mbox{and}\quad a=\frac{4+q}{q}\left(\frac{2}{q}\right)^{2/q},
\end{equation}
where $C$ is an integral constant. Therefore, we have two different methods to obtain the solution of the non linear differential equation, one gives a particular solution and the other one gives a parametric solution.
\section{Israel-Stewart Cosmological Model}
\begin{figure}[ht]
  \centering
  \includegraphics[height=5cm]{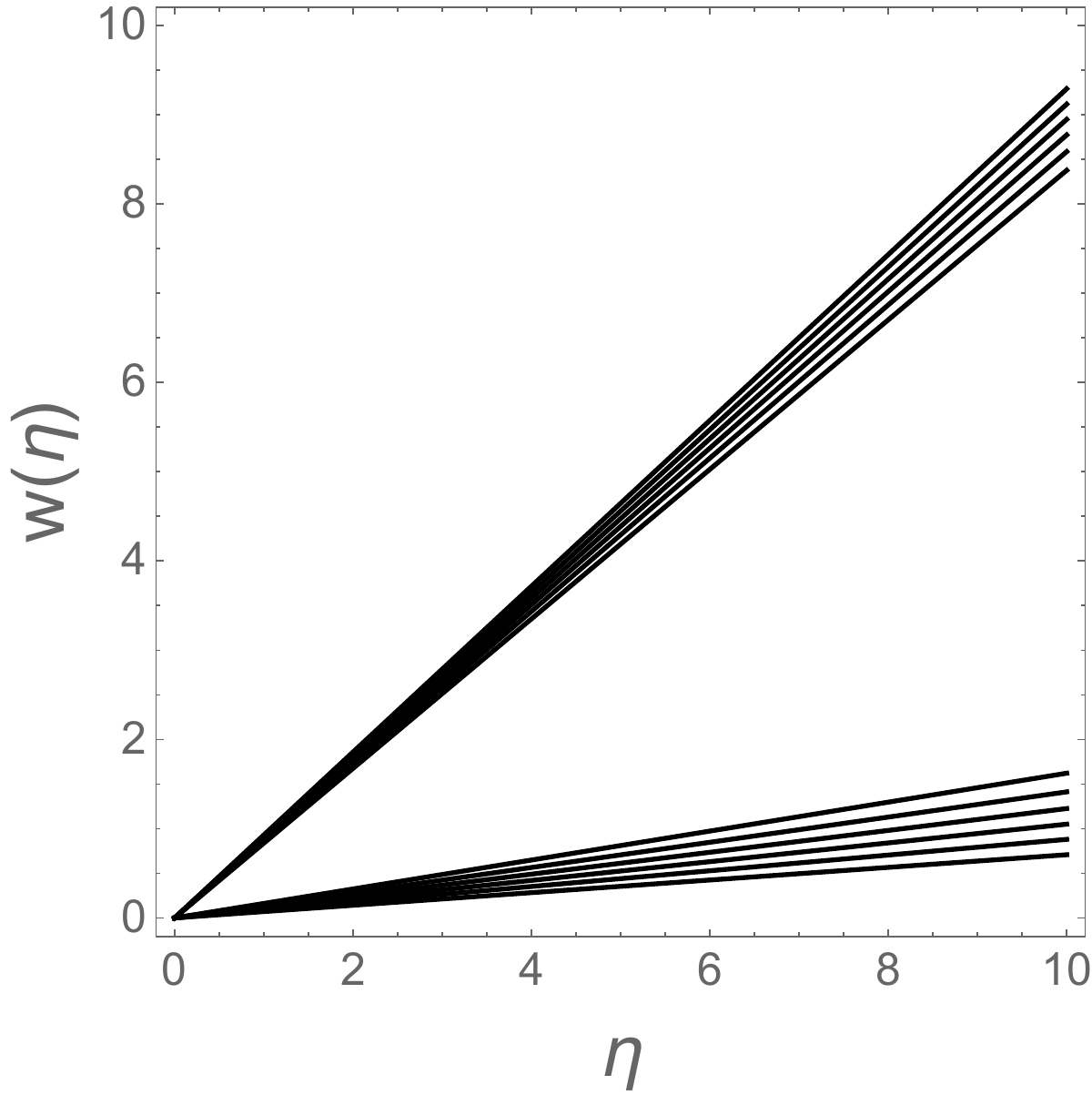}
   \caption{The figure shows the particular solution for the Hubble rate function given in terms of $w$ as a function of $\eta$ for different values of $0\leq\omega<1$. We have taken $\epsilon=\xi_0=3/4$.}
  \label{fig3}
\end{figure}

\begin{figure}[ht]
  \centering
  \includegraphics[height=5cm]{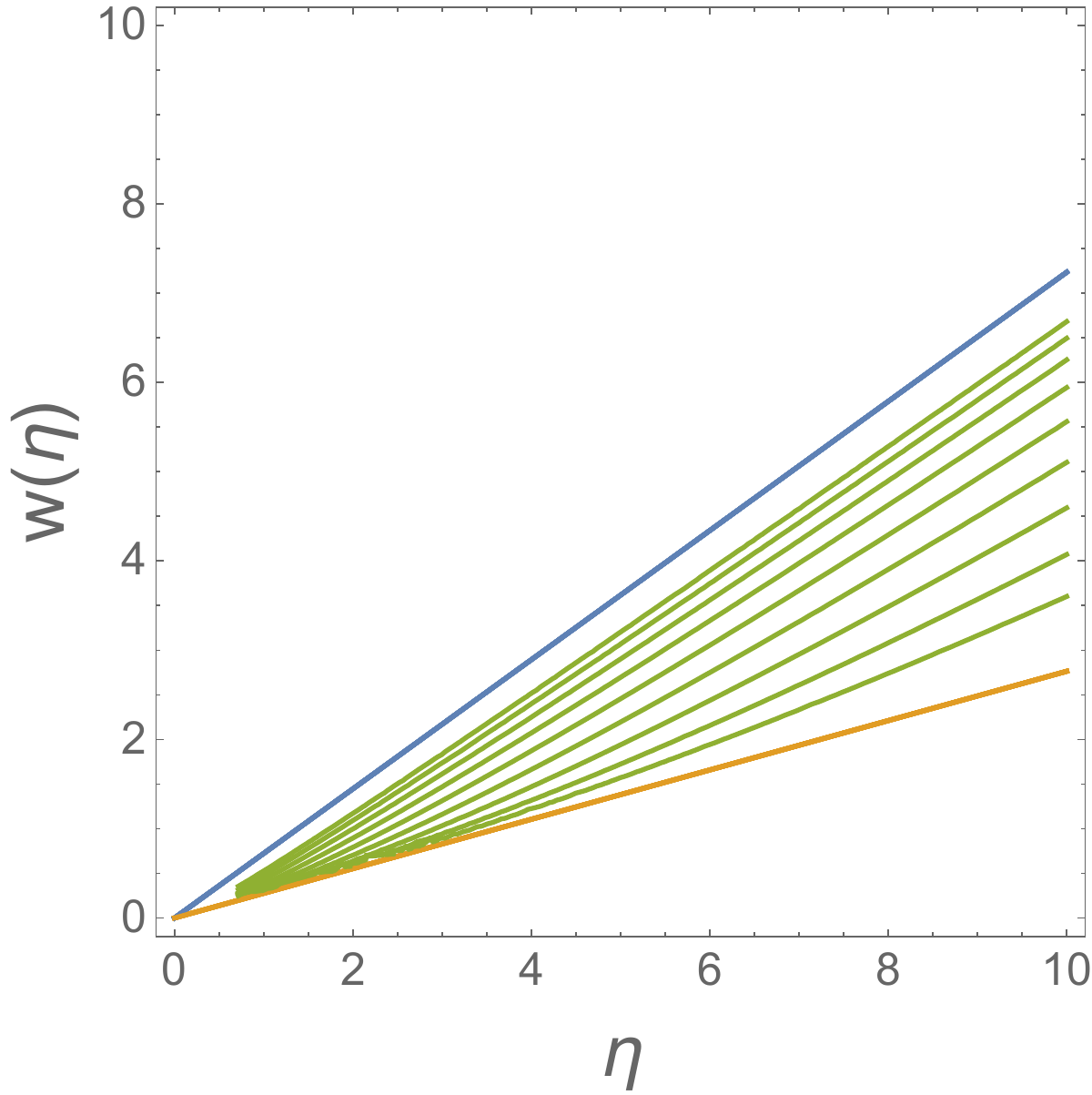}
   \caption{The figure shows the particular (blue and orange lines) and parametric solutions (green lines) given as a function of $\eta$ for the case when $\alpha_3(2+\alpha_1)/\alpha^2_2=1/5$ and $\epsilon=\xi_0=3/4$. For the parametric solution we have used different values of the integration constant $C$. The upper and lower lines correspond to the particular solution.}
  \label{fig4}
\end{figure}

A description of the relativistic thermodynamics of non-perfect fluids is given by the so called Israel-Stewart cosmological model. For the case when the bulk viscosity coefficient $\xi$ is given as a power law function of the energy density by $\xi=\xi_0\rho^{1/2}$ a cosmological solution of the polynomial type given by $H\propto(t+const.)^{-1}$ has been found by applying the standard factorization method \cite{cruz2019constraining,belinchon2022exact}, where $H$ denotes the Hubble rate function. The nonlinear differential equation for the Hubble function is given as the following mixed Li\'enard equation\cite{cruz2020exact}
\begin{equation}\label{eq20}
  \ddot{H}+\alpha_1\frac{\dot{H}^2}{H}+\alpha_2 H\dot{H}+\alpha_3 H^3=0
\end{equation}
where
\begin{align}\label{eq21}
   \alpha_1 =& -\frac{3}{2\delta} \\
  \alpha_2 = & \frac{3}{2}+3(1+\omega)-\frac{9}{4\delta}(1+\omega)+\frac{\sqrt{3}\epsilon(1-\omega^2)}{\xi_0} \\
  \alpha_3  =& \frac{9}{4}(1+\omega)+\frac{9}{2}\epsilon(1-\omega^2)\left[\frac{1+\omega}{\sqrt{3}\xi_0}-1\right]\\
  \delta(\omega) \equiv & \frac{3}{4}\left(\frac{1+\omega}{1/2+\omega}\right),
\end{align}
are constant coefficients and $0\leq\omega<1$. In Ref.\cite{cruz2019constraining}, Cruz {\it et al.} factorized equation (\ref{eq20}) in the following form
\begin{equation}\label{eq22}
  \left(D-\phi_1(H)\dot{H}-\phi_2(H)\right)\left(D-\phi_3(H)\right)H=0,
\end{equation}
where, after some algebra they found the following factorization
\begin{equation}\label{eq23}
  \left(D+\alpha_1\frac{\dot{H}}{H}-a^{-1}_1H\right)\left(D-a_1\alpha_3 H\right)H=0,
\end{equation}
where
\begin{equation}\label{eq24}
  a_1=\frac{-\alpha_2\pm\sqrt{\alpha^2_2-4\alpha_3(2+\alpha_1)}}{2\alpha_3(2+\alpha_1)}.
\end{equation}
Note that the factorization given in equation (\ref{eq22}) uses three functions and therefore it is more difficult to apply than the parametric factorization introduced in this paper. A particular solution is obtained after solving the first order differential equation given by
\begin{equation}\label{eq25}
  \dot{H}-a_1\alpha_3 H^2=0.
\end{equation}
In this section we are going to obtain the Hubble function by means of the parametric factorization. By assuming that the functions have the following form $\varphi_1=\tilde{a}_1\sqrt{(1+\alpha_1)\alpha_3}H$ and $\varphi_2=\tilde{a}^{-1}_1\sqrt{(1+\alpha_1)\alpha_3}H$ and using the parametric factorization conditions given in equations (\ref{eqc}) and (\ref{eqc1}) we obtain
\begin{align}\label{eq26}
  \varphi_1(H) =& (1+\alpha_1)\left[\frac{-\alpha_2\pm\sqrt{\alpha^2_2-4\alpha_3(2+\alpha_1)}}{2(2+\alpha_1)}\right] H\\
  \varphi_2(H) =&\frac{2\alpha_3(2+\alpha_1)}{-\alpha_2\pm\sqrt{\alpha^2_2-4\alpha_3(2+\alpha_1)}}H.
\end{align}
Therefore, equation (\ref{eq20}) admits the following parametric factorization
\begin{equation}\label{27}
  \left(D-\frac{2\alpha_3(2+\alpha_1)}{-\alpha_2\pm\sqrt{\alpha^2_2-4\alpha_3(2+\alpha_1)}}H\right)\left(D- (1+\alpha_1)\left(\frac{-\alpha_2\pm\sqrt{\alpha^2_2-4\alpha_3(2+\alpha_1)}}{2(2+\alpha_1)}\right) H\right)H^{1+\alpha_1}=0.
\end{equation}
To obtain a particular solution we have to solve the following first order differential equation
\begin{equation}\label{eq28}
  \left(D- (1+\alpha_1)\left(\frac{-\alpha_2\pm\sqrt{\alpha^2_2-4\alpha_3(2+\alpha_1)}}{2(2+\alpha_1)}\right) H\right)H^{1+\alpha_1}=H^{\alpha_1}(1+\alpha_1)\left[\dot{H}-a_1\alpha_3 H^2\right]=0,
\end{equation}
which has the same solution as the previous standard factorization given by
\begin{equation}\label{eq29}
  H(t)=\frac{A_{\pm}}{t-\left(t_0-\frac{A_{\pm}}{H_0}\right)},
\end{equation}
where $A_{\pm}=-(a_1\alpha_3)^{-1}$ and $H_0=H(t_0)$ is the Hubble constant.\\
We are now going to solve equation (\ref{eq20}) by changing it to an Abel equation as we did in the previous section. By using equation (\ref{eq16}) we arrive at the following Abel equation of the second kind
\begin{equation}\label{eq30}
  w\frac{dw}{dH}=\alpha_2H^{1+\alpha_1}w-\alpha_3H^{3+2\alpha_1}=\alpha_2 H^{1+\alpha_1}\left(w-\frac{\alpha_3}{\alpha_2}H^{2+\alpha_1}\right).
\end{equation}
By making the transformation $\eta=(\alpha_2/(2+\alpha_1))H^{2+\alpha_1}$, equation (\ref{eq30}) becomes the Abel equation in canonical form
\begin{equation}\label{eq31}
  w\frac{dw}{d\eta}=w-\alpha_3\left(\frac{2+\alpha_1}{\alpha^2_2}\right)\eta,
\end{equation}
which has a solution in parametric form given by\cite{zaitsev2002handbook}
\begin{equation}\label{32}
  \eta(\tau)=\frac{\alpha_2}{2+\alpha_1}H^{2+\alpha_1}(\tau)=C\exp\left(-\int\frac{\tau}{\tau^2-\tau-A}d\tau\right),\quad w(\tau)=C\tau\exp\left(-\int\frac{\tau}{\tau^2-\tau-A}d\tau\right),
\end{equation}
where $C$ is a constant and $A=\alpha_3(2+\alpha_1)/\alpha^2_2$. \\
It is interesting to  note that the particular and parametric solutions satisfy the following dynamic equation
\begin{equation}\label{eq33}
  \frac{dH^{1+\alpha_1}}{dt}=-\left(1+\alpha_1\right)w\left(H(t)\right)
\end{equation}
where
\begin{equation}\label{eq34}
  w(H)=-\frac{H^{1+\alpha_1}}{1+\alpha_1}\varphi_1(H)
\end{equation}
is the transformation used to obtain Abel´s equation. Therefore, using equation (\ref{eq33}) and comparing with equation (\ref{27}) we can express the particular solution as the following dynamic equation
\begin{equation}\label{eq35}
  \frac{dH^{1+\alpha_1}}{dt}=-\left(1+\alpha_1\right)\left(1\pm\sqrt{1-\frac{4\alpha_3(2+\alpha_1)}{\alpha^2_2}}\right)\eta.
\end{equation}
In figure (\ref{fig3}) we show the graph $w$ vs. $\eta$ for the particular solution for several values of $0\leq\omega<1$.\\
In figure (\ref{fig4}) we compare the particular and the parametric solutions where we have taken $\alpha_3(2+\alpha_1)/\alpha^2_2=1/5$. Note that the particular solutions are the upper and lower curves of the figure.\\
\section*{Conclusions}
In summary, compared with the factorization proposed by Rosu \& Cornejo-P\'erez the parametric factorization presented in this paper is more general and contains as a special case the standard factorization. The parametric factorization can deal with a larger class of non linear second order differential equations and also one can use the factorization conditions in order to obtain parametric solutions for a given non linear differential equation. Using the parametric factorization technique we have obtained a particular solution and a parametric solution of the generalized Fisher equation and the Israel-Stewart cosmological model in order to illustrate this approach. The parametric factorization presented in this paper can be used in other non linear mixed Li\'enard type equations.

\end{document}